\documentclass{mem}
\usepackage{natbib}\usepackage{txfonts}\usepackage{balance}
\usepackage{graphicx}
\usepackage[a4paper,breaklinks,dvipdfm]{hyperref}
\idline{75}{282}
\begin{document}
\def\teff{$T\rm_{eff }$}
\def\kms{$\mathrm {km s}^{-1}$}

\title{
VVV High Proper Motion Survey
}

   \subtitle{}

\author{
M. \,Gromadzki\inst{1}, \,R. \,Kurtev\inst{1}, \,S. \,Folkes\inst{1,2}, \,J. \,C. \,Beam{\'i}n\inst{3}, 
\,K. \,Pe\~{n}a Ram{\'i}rez\inst{4}, \\ 
J. \,Borissova\inst{1}, \,D. \,Pinfield\inst{2},  \,H. \,Jones\inst{2}, 
\,D. \,Minniti\inst{3,5,6} \and V. \,D. \,Ivanov\inst{7}
}

  \offprints{M. Gromadzki}

\institute{
Departamento de F{\'i}sica y Astronom{\'i}a, Facultad de Ciencias, Universidad de Valpara{\'i}so, 
Ave. Gran Breta\~{n}a 1111, Playa Ancha,Valpara{\'i}so, Chile
\email{mariusz.gromadzki@uv.cl}
\and
Centre for Astrophysics Research, Science and Technology Research Institute, 
University of Hertfordshire, Hatfield AL10 9AB, United Kingdom
\and
Instituto de Astrof\'isica, Facultad de F\'isica, Pontificia Universidad Cat\'olica de Chile,
Avda. Vicu\~na Mackenna 4860, 782-0436 Macul, Santiago, Chile\
\and
Instituto de Astrof{\'i}sica de Canarias, C/. V{\'i}a L{\'a}ctea s/n, E-38205 La Laguna, Tenerife, Spain
\and
The Milky Way Millennium Nucleus, Avda. Vicu\~{n}a Mackenna 4860, 782-0436 Macul, Santiago, Chile
\and
Vatican Observatory, Vatican City State V-00120, Italy
\and
European Southern Observatory, Avda. Alonso de Cordoba 3107, Casilla 19001, Santiago, Chile
}

\authorrunning{M. Gromadzki}

\titlerunning{VVV High Proper Motion Survey}

\abstract{Here we present survey of proper motion stars towards the Galactic Bulge and an adjacent plane region base on VISTA-VVV data. 
The searching method based on cross-matching photometric $K_{\rm S}$-band CASU catalogs. The most interesting discoveries are shown. 

}
\maketitle{}

\section{Introduction}

High proper motion searches is the most widespread method of revealing new solar neighbors, as nearby stars generally have larger proper motions than those more distant. The Galactic bulge and plane are referred as the zone of avoidance, because they contain very high stellar densities down to faint limiting magnitudes, and also regions of dark molecular clouds, nebulosity, as well as regions of current star formation. Most of nearby stars are found outside these regions. Although, recent discoveries of nearby UCDs in Galactic plane, DENIS J081730.0-615520 \citep{Artigau}, UGPS J072227.51-054031.2 \citep{Lucas} and Luhman 16 \citep{Luhman}, shows that these regions may contain other highly interesting, unusual, and nearby objects.

\section{VISTA-VVV}

The ESO public survey VISTA Variables in the V{\'i}a L{\'a}ctea (VVV) targets 562 square degrese in the Galactic Bulge and an adjacent plane region \citep{Minniti,Saito}. VVV provides multi-epoch $K_{\rm S}$-band images which allow searching for high proper motion objects. Providing better spatial resolution and deeper (∼4) magnitude range than 2MASS, VVV has bigger potential of finding free floating very low-mass stars, brown dwarfs and also common proper motion companions of previously known high proper motion stars. The VVV data are processed by the Cambridge Astronomical Unit (CASU)\footnote{http://casu.ast.cam.ac.uk/vistasp/}. CASU provide images and catalogs with aperture photometry and astrometry for VVV users. The VVV data are also publicly available through the VISTA Science Archive (VSA)\footnote{http://horus.roe.ac.uk/vsa/index.html}. More technical information about the VVV survey can be found in \citep{Saito} and \citep{Soto}.

\begin{figure*}[t]
\resizebox{\hsize}{!}{\includegraphics[clip=true]{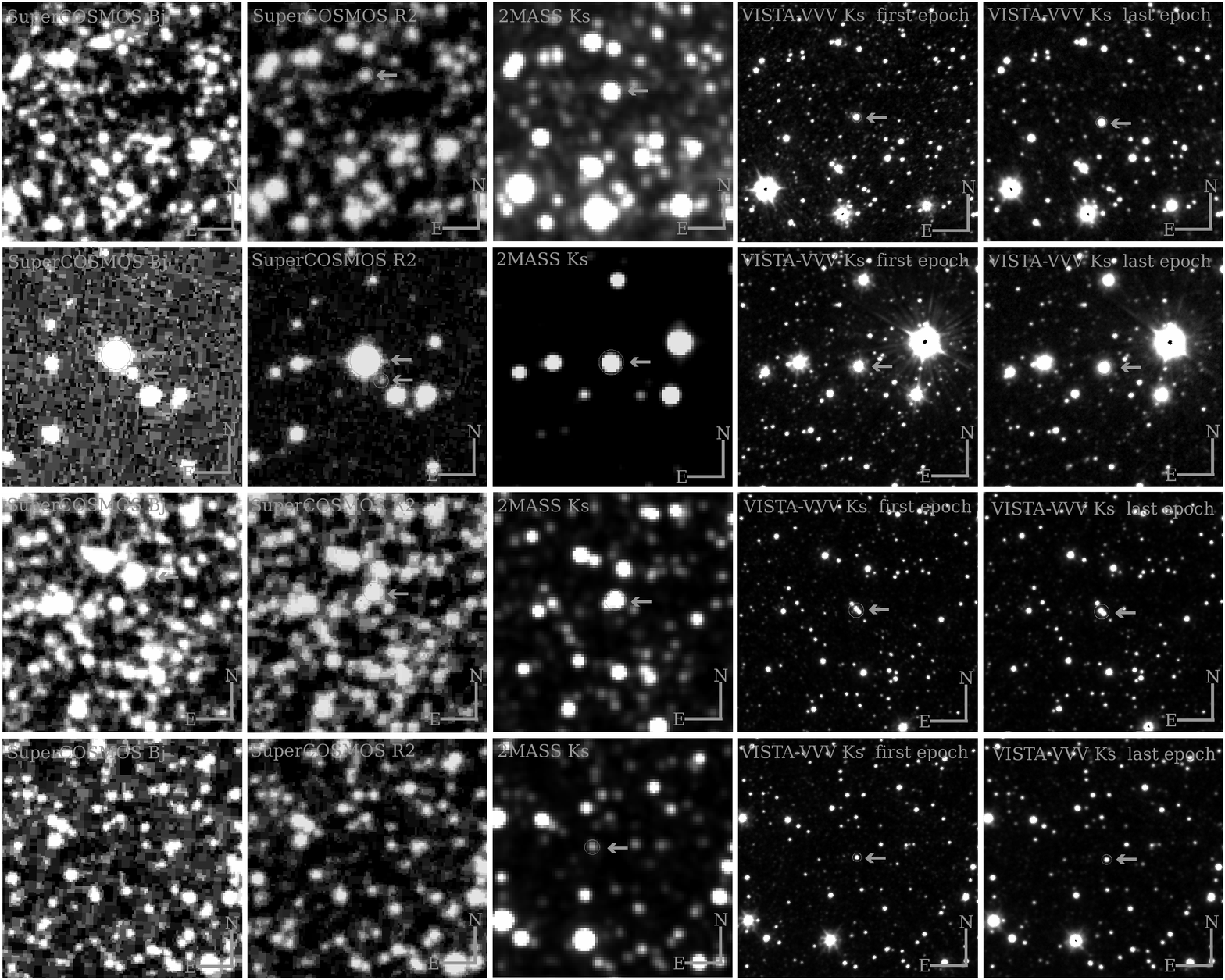}}
\caption{\footnotesize
Examples of most interesting discoveries.
Row 1: High proper motion (0.81''/yr) M dwarf towards the Galactic Bulge. Row 2: MD+WD common proper motion binary, WD is visible only on SuperCOSMOS images  ($B_{j}$=20.3, $R_{2}$=21.1). Row 3: Close common proper motion MD+MD binary ($d$=1.6'') resolved on VISTA-VVV images. Row 4: Brown dwarf towards the Galactic Bulge, object is not visible on SuperCOSMOS images.
}
\label{eta}
\end{figure*}

\section{Method}

The searching method we used is based on cross-matching, with a scaled radius of 5 arcsec per year, photometric $K_{\rm S}$-band CASU catalogs obtained for 4 different epochs. Method was developed by Folkes et al. (in preparation). Next, the candidates were visually inspected on VVV, 2MASS and SuperCOSMOS images.

\section{Current status}

Until Jun 2013, we have examined ∼31 \% of VVV area and we have detected ∼400 objects with proper motion higher than 0.05 arcsec per year and $K_{\rm S}<$13.5. This sample includes dozens of completely new high proper motion stars and common proper motion pairs, some common proper motions companions of previously known high proper motion stars, and one spectroscopically confirmed brown dwarf (Beam{\'i}n et al., submitted to A\&A). During visual inspection of SuperCOSMOS images we have also identified 3 white dwarf common proper motion companions of previously known high  proper motion stars.

The examples of most interesting discoveries are presented on Figure 1. SuperCOSMOS $B_{j}$ ans $R_{2}$, 2MASS $K_{\rm S}$ and VISTA-VVV first and last $K_{\rm S}$ epochs images were obtained aroud 1974, 1992, 2000, 2010 and 2012, respectively.

\begin{acknowledgements}
We gratefully acknowledge use of data from the ESO Public Survey programme ID 179.B-2002 taken with the VISTA telescope, data products from the Cambridge Astronomical Survey Unit, and funding from the BASAL CATA Center for Astrophysics and Associated Technologies PFB-06. This publication makes use of data products from the Two Micron All Sky Survey,  which is a joint project of the University of Massachusetts and the Infrared Processing 
and Analysis Center, funded by the National Aeronautics and Space Administration and the National Science Foundation. Our research has also made use of data obtained from the SuperCOSMOS Science Archive, prepared and hosted by the Wide Field Astronomy Unit, Institute for Astronomy, University of Edinburgh, which is funded by the UK Science and 
Technology Facilities Council. MG is financed by the GEMINI-CONICYT Fund, allocated to Project 32110014. RK and JB acknowledge partial support from FONDECYT through grants No 1130140 and 1120601 respectively. J.C.B. acknowledges support from a PhD Fellowship from CONICYT. D.M. acknowledges support by Project FONDECYT No. 1130196. MG is also grateful to organiser of ''Brown Dwarfs come of Age'' conference for partial cover expanses of his participation in the conference.  
\end{acknowledgements}

\bibliographystyle{aa}

\begin{thebibliography}{}


\bibitem[Artigau et al.(2010)]{Artigau} Artigau, {\'E}., 
Radigan, J., Folkes, S., et al.\ 2010, \apjl, 718, L38 


\bibitem[Lucas et al.(2010)]{Lucas} Lucas, P.~W., Tinney, 
C.~G., Burningham, B., et al.\ 2010, \mnras, 408, L56 

\bibitem[Luhman(2013)]{Luhman} Luhman, K.~L.\ 2013, \apjl, 
767, L1 


\bibitem[{Minniti et al. (2010)}]{Minniti} 
Minniti, D., Lucas, P.W., Emerson, J.P. et al. 2010, New Astronomy, 15, 433

\bibitem[Saito et al.(2012)]{Saito} Saito, R.~K., Hempel, M., Minniti, D., et al.\ 2012, \aap, 537, A107 

\bibitem[Soto et al.(2013)]{Soto} Soto, M., Barb{\'a}, R., Gunthardt, G., et al.\ 2013, \aap, 552, A101


\end{thebibliography}

\end{document}